# Anisotropic Thermal Conductivity Measurement of Organic Thin Film with Bidirectional 3$\omega$ Method

*Shingi Yamaguchi[1], Takuma Shiga[1], Shun Ishioka[2], Tsuguyuki Saito[2], Takashi Kodama[1], Junichiro Shiomi[1] \**

1. Department of Mechanical Engineering, The University of Tokyo, 7-3-1, Hongo, Bunkyo-ku, Tokyo 113-8656

2. Department of Biomaterial Sciences, Graduate School of Agricultural and Life Sciences, The University of Tokyo, 1-1-1 Yayoi, Bunkyo-ku, Tokyo 113-8657

*shiomi@photon.t.u-tokyo.ac.jp

ABSTRACT

Organic thin film materials with molecular ordering are gaining attention as they exhibit semiconductor characteristics. When using them for electronics, the thermal management becomes important, where heat dissipation is directional owning to the anisotropic thermal conductivity arising from the molecular ordering. However, it is difficult to evaluate the anisotropy by simultaneously measuring in-plane and cross-plane thermal conductivities of the film on a substrate, because the film is typically as thin as tens to hundreds of nanometers and its in-plane thermal conductivity is low. Here, we develop a novel bidirectional 3ω system that measures the anisotropic thermal conductivity of thin films by patterning two metal wires with different widths and preparing the films on top, and extracting the in-plane and cross-plane thermal conductivities using the difference in their sensitivities to the metal-wire width. Using the developed system, the thermal conductivity of spin-coated poly(3,4-ethylenedioxythiophene) polystyrene sulfonate (PEDOT:PSS) with thickness of 70 nm was successfully measured. The measured in-plane thermal conductivity of PEDOT:PSS film was as high as 2.9 W m$^{-1}$ K$^{-1}$ presumably due to the high structural ordering, giving anisotropy of 10. The calculations of measurement sensitivity to the film thickness and thermal conductivities suggest that the device can be applied to much thinner films by utilizing metal wires with smaller width.

## INTRODUCTION

In the field of electronics, organic thin films are gaining attention because of their semiconductor features[1,2]. Thermal management of these materials is an important issue as the heat is generated due to self-Joule heating, and therefore the evaluation of their thermal conductivity ($\kappa$) is essential. Many of these materials take highly ordered structure in the form of thin film with a thickness less than 100 nm[3,4], and this structural anisotropy and small thickness have to be considered in the $\kappa$ measurements. Among the developed measurement techniques for in-plane thermal conductivity ($\kappa_\parallel$) such as the off-set time domain thermoreflectance (TDTR) method[5–7], suspended microdevice[8,9], and off-set laser flash method[10], 3ω method[11–16] has an advantage in the capability to measure such thin samples. The 3ω method separately measures cross-plane thermal conductivity ($\kappa_\perp$) and thermal anisotropy ($\eta = \kappa_\parallel/\kappa_\perp$) by detecting the voltage oscillation signal with various widths of metal wires deposited on the sample. The measurement is made sensitive to $\eta$ by making the width of a narrowest wire as small as the thickness of thin film. Such metal wires are usually prepared on the target sample by lithography techniques, which can damage organic materials with harsh chemicals in the developing or etching process. Some alternative methods have been proposed for the $\kappa_\parallel$ measurement of organic thin films, by fabricating metal wires with shadow mask[17,18] or applying 3ω measurement to a film on a suspended membrane with metal wires[19,20]. However, these methods also have limits on appreciable sample range; the shadow mask cannot prepare nanometers-wide wire required for the measurement of submicron-thin film, and the suspended membrane can be destructed by the common spin coating process used for film preparation. Therefore, a general method with wider applicability is required to measure $\eta$ of organic thin films.

There also exists the bidirectional 3ω method, where a sample is placed on metal wires which is prepared beforehand on a substrate[21–24]. Here, "bidirectional" is named after the directions of heat conduction, in comparison to the conventional 3ω method where the heat conducts unidirectionally (Figure 1a,b). While this method can avoid damaging the sample by chemicals, only single narrow wire has been used to extract $\kappa_\perp$ and $\eta$, since the measurement with wider wire would be insensitive to both $\kappa_\perp$ and $\eta$[25]. In this work, we develop a bidirectional 3ω system with wide and narrow wires to measure the anisotropic thermal conductivities of organic thin films. In analogy to the conventional 3ω system (Figure 1a), $\kappa_\perp$ was obtained first from measurement with wide wire, where thermally conductive Al layer was deposited to selectively suppress the sensitivity to $\eta$ (Figure 1b), and then $\eta$ was obtained from measurement with narrow wire without the Al layer. By preparing the wires with different widths on one substrate, the two parameters ($\kappa_\perp$ and $\eta$) can be obtained from the same sample. The method can measure the thermal anisotropy of thin organic film with a thickness of less than 100 nm, and the results clarify that $\kappa_\parallel$ of highly ordered organic thin film can be more than three times larger than ever reported.

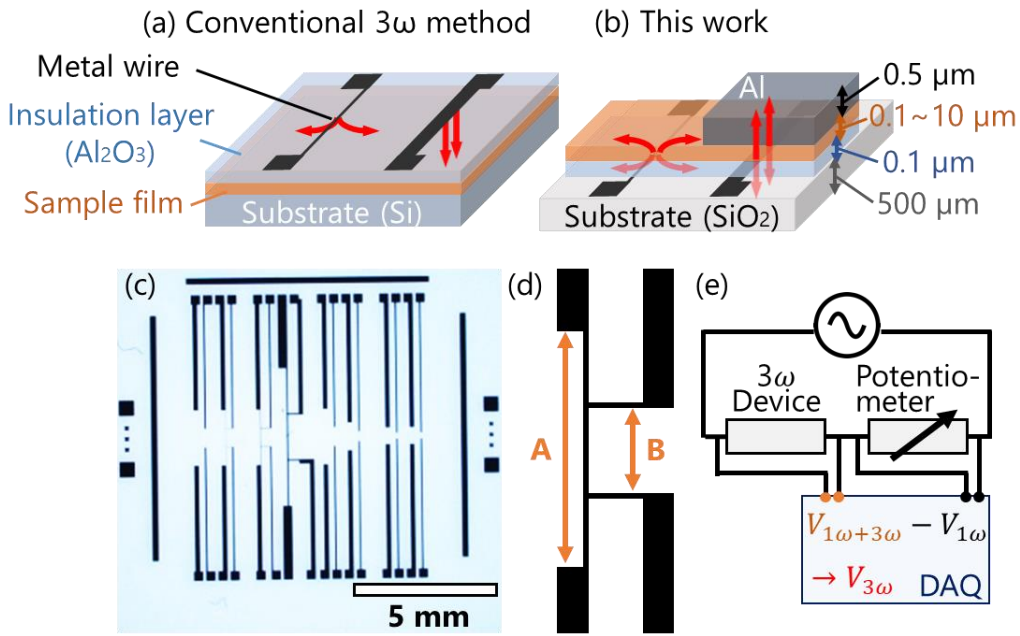

Figure 1 Schematic image of (a) conventional 3ω method and (b) bidirectional 3ω method in this work. (c) Picture of the measurement device. (d) Enlarged image of the metal wire. (e) Schematic image of measurement setup.

EXPERIMENTAL METHOD

・Device and sample preparation: The measurement device was designed to have 8 metal wires of Ti / Pt with different widths (Figure 1c). The wire width and the lengths between current probes and between voltage probes (length A and B in Figure 1d, respectively) are summarized in Table 1. Thirty devices were fabricated on the 4-inch quartz wafer at once and all the processes were conducted in Takeda cleanroom at the University of Tokyo. Quartz was chosen as the substrate material to suppress heat conduction to the substrate. Ti / Pt (5- and 20-nm-thick, respectively) was sputtered on the substrate first, and negative resist for electron beam lithography was spin-coated. The patterns were prepared by direct exposure, development, and subsequent dry etching. A 100 nm alumina ($Al_2O_3$) layer was deposited on the device as an insulation layer by atomic layer deposition (ALD), while $Al_2O_3$ on contact pads was removed by wet-etching process using the photoresist and mask aligner. The wafer was cut into square-shaped devices with the size of 1 $cm^2$. As the wire with 100-nm width did not survive the dry etching process, the 300-nm-wide wire was used as the narrowest one in this study. Polymethyl methacrylate (PMMA) and PEDOT:PSS films were prepared by spin coating, and part of the film on the contact pads was wiped off before the measurement. The details of sample preparation are provided in Supplementary Material.

Table 1 Length between current and voltage probes for each wire width.

| Wire width / μm | 50    | 20    | 10    | 2     | 1     | 0.5 | 0.3 |
|---|---|---|---|---|---|---|---|
| Length A / μm   | 4,500 | 1,800 | 1,800 | 1,800 | 1,800 | 900 | 540 |
| Length B / μm   | 1,500 | 600   | 600   | 600   | 600   | 300 | 180 |

・Solution to the bidirectional heat diffusion equation: The two-dimensional equation of Feldman's algorithm[26] was chosen to take the thermal anisotropy into account. The calculated temperature rise ($\Delta T$) at the wire can be written as:

$$\Delta T = \frac{P}{2\pi l}\int_0^\infty \frac{(A^+ + A^-)(B^+ + B^-)}{A^+B^- - A^-B^+}\frac{1}{\gamma_j}\frac{\sin^2(kb)}{(kb)^2}dk \quad \cdots(1)$$

$$\gamma_j = \kappa_{j,z}\sqrt{\eta_j k^2 + \frac{2i\omega}{D_{j,z}}} \quad \cdots(2)$$

where $P$, $l$, and $b$ are the heating power, length, and half width of the wire, respectively. $\kappa_{j,z}$ and $D_{j,z}$ are the thermal conductivity and thermal diffusivity of the jth layer in the z direction, and $\eta_j$ is anisotropy ($\kappa_{j,x}/\kappa_{j,z}$). $k$ is a variable of integration, and $A^+$, $A^-$, $B^+$ and $B^-$ are dimensionless parameters obtained by solving the equation by a recursive matrix method. The details of mathematical solution and material parameters required for calculation (specific heat capacity and mass density) are included in Supplementary Material.

The absolute value of measurement sensitivities ($|S_\beta|$) to the parameters of interest ($\beta$) can be calculated using Eq. (1) as:

$$|S_\beta| = \left|\left(\frac{\Delta\beta}{\beta}\right)^{-1}\frac{\Delta T(\beta + \Delta\beta) - \Delta T(\beta)}{\Delta T(\beta)}\right| \quad \cdots(3)$$

where $\Delta\beta/\beta$ was set to be 0.1 referring to the previous work[21].

・Measurement setup and analysis procedure: For the measurement, the metal wire on the 3ω device was connected in series with a potentiometer. A schematic of the setup is shown in Figure 1e. The potentiometer is adjusted to have the same electrical resistance with that of the wire ($R$). The temperature coefficient of the resistance ($\alpha$) of the wire was obtained in advance by measuring the $R$ from 20 to 40 °C with increment of 5 °C. All the measurement was conducted under room temperature (25 °C). Wave current with a frequency $\omega$ (10 ~ 1500 Hz) was applied, and the third harmonic voltage across the wire ($V_{3\omega}$) was obtained by subtracting the first harmonic voltage across the potentiometer ($V_{1\omega}$) from that across the device ($V_{1\omega+3\omega}$). The voltage was measured by DAQ NI-9239 (National Instruments). The temperature rise ($\Delta T$) at the wire was calculated by

$$\Delta T = \frac{2}{\alpha} \frac{V_{3\omega}}{V_{1\omega}} \quad \cdots (4)$$

and the current amplitude ($I_{1\omega}$) was controlled so that the highest $\Delta T$ is around 2 K. The target parameters were obtained by fitting the theoretical $\Delta T$ (Eq. (1)) with the experimental ones (Eq. (4)) with a least squares method.

The measurement was firstly conducted with bare device, and the $\kappa$ of insulation layer ($Al_2O_3$) was determined for differential measurement. Next, measurements with (i) the film and (ii) the film with deposited Al on the top were conducted to get $\Delta T$. Finally, $\kappa_\perp$ and $\eta$ of the film were determined in order by fitting the data of (ii) and (i), respectively.

・Sources of error: In this study, measurements on a target material were conducted 3 times with different substrates, and a standard deviation of the obtained values is denoted as the error. Possible sources of the deviation are (1) the fluctuation of the measured signal and (2) the geometric uncertainty of the measurement setup. As for (1), in our system, DAQ functioning as a lock in amplifier can detect $V_{3\omega}$ signal with fluctuation as low as 0.2 %, which is negligible. Regarding (2), the uncertainties of the wire width and the layer thickness influence the fitting results. As will be discussed later, differential measurements were conducted when obtaining $\eta$ to eliminate the uncertainty of Al$_2$O$_3$ layer thickness, but those of the wire width and the sample thickness remain. Their impact on the target properties can be estimated by inputting the geometries altered by 5 % into the physical model to simulate the signal and then fitting it with the physical model assuming unaltered geometries. For example, with the PMMA film case, the deviation was estimated to be 8 % in the cross-plane thermal conductivity and 29 % in the thermal anisotropy, which reasonably match with the above standard deviation obtained from multiple experiments.

RESULTS AND DISCUSSION

Before measuring the organic thin films, thermal conductivity of the 100-nm-thick insulation layer $\kappa_{Al_2O_3}$ was determined. Figure 2a shows the measurement sensitivities to $\kappa_{Al_2O_3}$ with wide (50 µm) and narrow (300 nm) wires, which were calculated assuming that $\kappa_{Al_2O_3}$ is the reported value [27] of 2 W m$^{-1}$ K$^{-1}$ and thermal anisotropy $\eta_{Al_2O_3}$ is 1. Sensitivities to $\kappa_{SiO_2}$ were also shown as a reference. While 50-µm-wide wire has sensitivity only to $\kappa_{SiO_2}$ and not to $\kappa_{Al_2O_3}$ in the whole frequency range, the 300-nm-wide wire gives the sensitivity to $\kappa_{Al_2O_3}$ as high as 0.2. Therefore, the 300-nm-wide wire was used for the measurement of $\kappa_{Al_2O_3}$. The reason for this sensitivity difference with wide- and narrow-wires are discussed in Supplementary Material. As input parameters, measured values were used for thickness and specific heat capacity, while reference values were used for density. The actual values of the parameters are summarized in Table S1. The theoretical curve calculated by Eq. (1) was fitted with experimental values with a least squares method by altering $\kappa_{Al_2O_3}$ as a fitting parameter. Figure 2b shows the experimental temperature rise and best fitted curve for the 300-nm-wide wire. The theoretical curve shows good agreement with the experimental data, and the fitting gives $\kappa_{Al_2O_3}$ of 1.50±0.72 W m$^{-1}$ K$^{-1}$ (N=6), which is consistent with the reported value[27]. The relatively large error may be due to the uncertainty in the device configuration, such as wire width and Al$_2$O$_3$ thickness, which deviate depending on the device position on the wafer. This uncertainty is eliminated in the measurement of the organic films by the differential technique as described later.

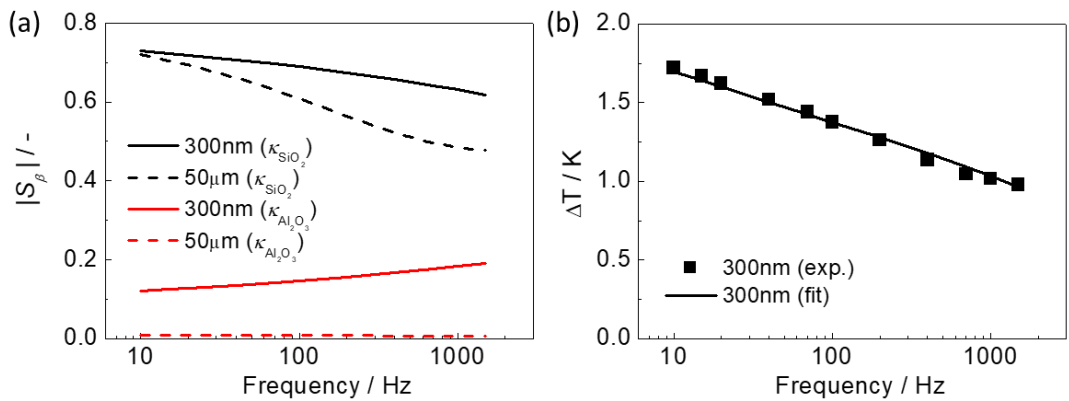

Figure 2 (a) Sensitivities to $\kappa_{Al_2O_3}$ and $\kappa_{SiO_2}$ when 300-nm- and 50-μm-wide wires are used. (b) Temperature rise and the fitting curve for the measurement of $\kappa_{Al_2O_3}$ with the 300-nm-wide wire.

Next, anisotropic $\kappa$ of cellulose nanofiber (CNF) oriented film was measured to validate the system. The oriented CNF film was selected as a model sample because its anisotropic $\kappa$ has been already reported[28], and also it can be formed either into a film on substrate or into a self-standing film, which makes it is a suitable sample for validation that can be measured by both bidirectional 3ω method and other existing methods. The CNF film thickness measured by micrometer was 26 μm. To calculate the measurement sensitivities, reported values of CNF cross-plane thermal conductivity ($\kappa_{CNF,\perp}$) and its thermal anisotropy ($\eta_{CNF}$) were used[28]. For thermal boundary resistance (TBR) values at all organic/inorganic interfaces, a reported value of $3\times10^{-8}$ $m^2\,K\,W^{-1}$ at the interface between silicon (Si) and spin-coated organic polymer measured by TDTR method[29] was employed. The absolute value of TBR is not likely to affect the results as the measurement sensitivity to TBR is less than 0.003 (See Figure S1). Sensitivities to $\kappa_{CNF,\perp}$ and $\eta_{CNF}$ with the 300-nm-wide wires are shown in Figure 3a. The sensitivities to $\kappa_{CNF,\perp}$ and $\eta_{CNF}$ are as high as 0.1 to 0.2, and another independent measurement is required to suppress uncertainty by determining either one of the parameters.

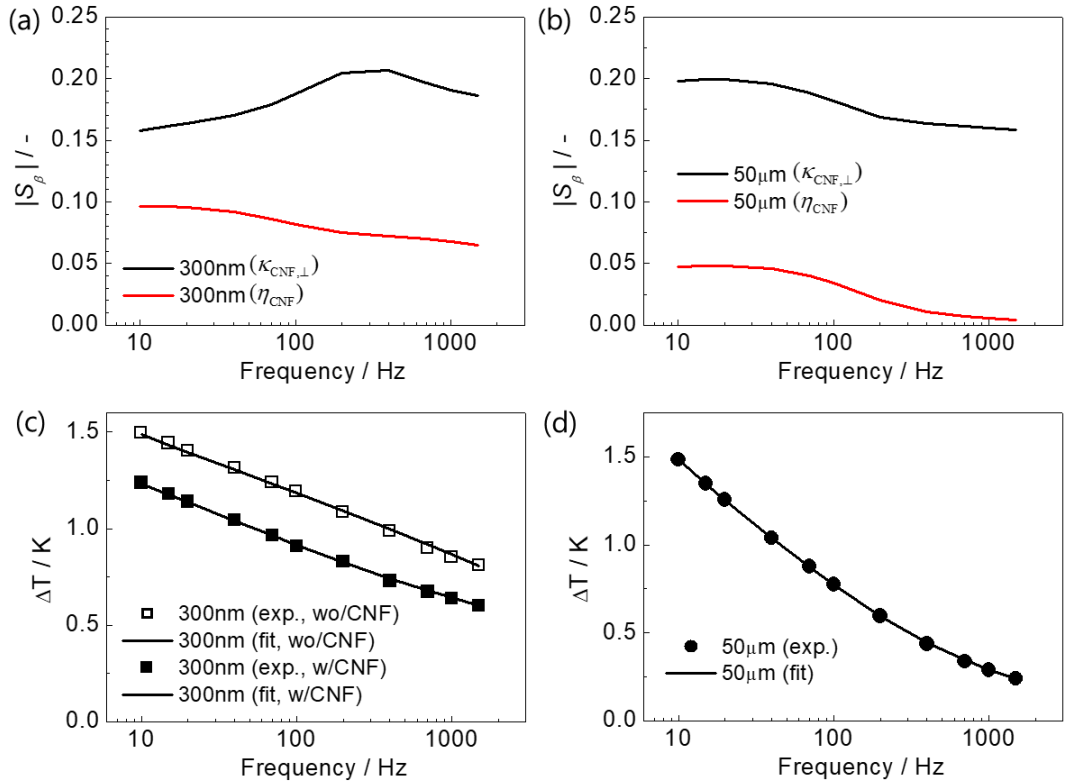

Figure 3 Sensitivities to $\kappa_{CNF,\perp}$ and $\eta_{CNF}$ when (a) the 300-nm-wide wire is used without Al deposition and (b) the 50-μm-wide wire is used with Al deposition. Representative temperature rise and its fitting curve for measurement of $\kappa_{CNF,\perp}$ and $\eta_{CNF}$ with (c) the 300-nm-wide wire and (d) the 50-μm-wide wire. Control measurement without CNF is also shown in (c).

In conventional 3ω measurement, the films are sandwiched by a metal wire and heat conducting substrate such as Si, and the generated heat passes through the film to the substrate (Figure 1a). In analogy to the conventional system, 500 nm of heat conducting aluminum (Al) was deposited on the CNF film with resistive thermal evaporation, so that the heat can conduct in the vertical direction (Figure 1b). The temperature rise of the sample during deposition is less than several tens of degrees as melting point of Al is low[30], and the heating effect to the sample is negligible. Al was also deposited on a bare device simultaneously to measure $\kappa$ of Al ($\kappa_{Al}$) in advance (the representative temperature rise and its fitting curve to extract $\kappa_{Al}$ is shown in Figure S2). Figure 3b shows sensitivity to each parameter when the 50-μm-wide wire is used to measure CNF with the Al deposition. On the contrary to the previous 300-nm case, the sensitivity to $\eta_{CNF}$ is less than 25 % of that to $\kappa_{CNF,\perp}$, which means that the measurement is far sensitive to $\kappa_{CNF,\perp}$. Therefore, $\kappa_{CNF,\perp}$ was firstly determined with the 50-μm-wide wire with the Al deposition, and then $\eta_{CNF}$ was determined with the 300-nm-wide wire without the Al deposition. To get those parameters from an identical sample, the measurements of the temperature rises were conducted in reverse order; firstly without Al case and secondly with Al case.

$\kappa_{CNF,\perp}$ and $\eta_{CNF}$ were obtained by the process described above. In addition to the thickness, specific heat capacity, and density (actual values summarized in Table S1), the reference values[28] of $\kappa_{CNF,\perp}$ and $\eta_{CNF}$ were used as initial input parameters, while measured value of $\kappa_{CNF,\perp}$ was used when obtaining $\eta_{CNF}$. A least squares method was used for fitting by altering $\kappa_{CNF,\perp}$ or $\eta_{CNF}$ as a fitting parameter in each case. Figure 3c,d shows the measured temperature rise and its fitting curve of each case for CNF measurement. $\kappa_{CNF,\perp}$ was determined to be 0.28±0.02 W m$^{-1}$

K$^{-1}$ (N=3) from the measurement with the 50-μm-wide wire (Figure 3d). A differential technique was used for the measurement of $\eta_{CNF}$; $\kappa_{Al_2O_3}$ was measured beforehand and the value was used in the measurement of $\eta_{CNF}$. This method can eliminate the uncertainty resulting from the device configuration as it is included in $\kappa_{Al_2O_3}$. Figure 3c shows the temperature rise of the wire with and without CNF, where the difference results from the increased heat conduction to CNF. The measured $\eta_{CNF}$ was 3.46±0.15 (N=3), and correspondingly the $\kappa_\parallel$ of CNF ($\kappa_{CNF, \parallel}$) was 0.95±0.04 W m$^{-1}$ K$^{-1}$. To validate these values, $\kappa_{CNF,\perp}$ and $\kappa_{CNF, \parallel}$ of self-standing CNF film were also measured by laser flash analysis (LFA)[31,32] and T-type method[33–35], respectively. The details of each method are described in Supplementary Material. $\kappa_{CNF,\perp}$ obtained from LFA was 0.30±0.03 W m$^{-1}$ K$^{-1}$, and $\kappa_{CNF, \parallel}$ obtained from T-type method was 1.09±0.07 W m$^{-1}$ K$^{-1}$. These values agree well with the ones obtained from bidirectional 3ω method, which proves the reliability of this novel method.

Thermal conductivities of spin-casted polymer films were also measured to reveal the anisotropic thermal properties of submicron-thin films. PMMA and PEDOT:PSS were selected as representative materials, and their films were spin-coated on the measurement device. The thickness of PMMA and PEDOT:PSS film were 308 nm and 70 nm (measured by interferometry and AFM, respectively). Figure S3 shows measurement sensitivities calculated with the reported cross-plane thermal conductivity and thermal anisotropy of PMMA[17] and PEDOT:PSS[36]. As is the same with CNF case, the sensitivities to both $\kappa_\perp$ and $\eta$ are large when narrow-wire was used, whereas sensitivities to $\eta$ is negligible when wide wire is used with Al deposition. Therefore, the same process was conducted to obtain $\kappa_\perp$ and $\eta$ of each materials (input parameters summarized in Table S1). The measured temperature rise and fitting curves are shown in Figure S4, and obtained $\kappa_\perp$, $\eta$, and $\kappa_\parallel$ are shown in Table 2 together with reference values. While $\kappa_\perp$ of PMMA and PEDOT:PSS are close to the reported ones, their $\kappa_\parallel$ are much different from the reference values. The lower value of the measured $\kappa_\parallel$ of PMMA could be due to the difference in the degree of polymer alignment. The PMMA film in this study has larger thickness (308 nm) than the reported one (170 nm), presumably because of the lower rotation speed in the spin-coat process, which might have led to the lower degree of alignment and hence the lower thermal conductivity[37,38]. On the other hand, $\kappa_\parallel$ of PEDOT:PSS is three to ten times higher than the reported values obtained from thicker films. This can be also explained by the difference of molecular ordering as the thicker films were formed by drop-casting or repeating spin-coatings, and therefore, 70-nm-thin PEDOT:PSS film is suggested to have high $\kappa_\parallel$ because of its highly ordered structure.

Table 2. Anisotropic thermal properties of spin-casted polymer films (N=3).

|  |  | $\kappa_\perp$ (W m$^{-1}$ K$^{-1}$) | $\kappa_\parallel$ (W m$^{-1}$ K$^{-1}$) | $\eta$ (= $\kappa_\parallel/\kappa_\perp$) |
|---|---|---|---|---|
| **PMMA** | This study | 0.22±0.01 | 0.72±0.18 | 3.25±0.83 |
| | Ref. | 0.20 ~ 0.25 [17,29,39–41] | 2.33±1.55 [17] | - |
| **PEDOT:PSS** | This study | 0.29±0.11 | 2.94±0.81 | 10.1±2.8 |
| | Ref. | 0.2 ~ 0.3 [36,42,43] | 0.3 ~ 0.8 [20,36,42,43] | - |

To examine the range of the sample properties and geometries that the current method is applicable to, sensitivities to $\kappa_\perp$ and $\kappa_\parallel$ of model sample with various thicknesses ($d$) were calculated. The density and specific heat capacity of model sample were set to 1000 J/kg·K and 1000 kg/m$^3$ respectively, as representative values of organic materials. The sensitivity of 0.02 was taken as lower bound considering that $\kappa_{\text{PEDOT:PSS}, \parallel}$ can be determined with 30 % uncertainty when the sensitivity is 0.02 (Figure S4). Firstly, sensitivity to $\kappa_\perp$ of model sample was calculated when the 50-μm-wide wire was used with 1000 Hz alternating current and 500-nm-thick Al is deposited on the sample. An isotropic model sample was assumed in this analysis to show the lowest bound, while actual sample is expected to have thermal anisotropy. The contour of 0.02 in Figure 4a indicates the lower bound of the measurement, which corresponds to a thermal resistance (=$d/\kappa_\perp$) of ~10$^{-7}$ m$^2$ K W$^{-1}$. The measurable smallest film thickness increases from 20 nm to 800 nm as the $\kappa_\perp$ increases from 0.1 W m$^{-1}$ K$^{-1}$ to 10 W m$^{-1}$ K$^{-1}$. This lower bound of thickness can become high if the sample has a thermal anisotropy. Next, sensitivity to $\kappa_\parallel$ of a model sample was estimated when the 300-nm-wide wire was used with 1000 Hz alternating current, without Al deposition. Here, $\kappa_\perp$ of the sample is assumed to be 0.1 W m$^{-1}$ K$^{-1}$, and the thermal anisotropy was changed from 1 to 100. The contour of 0.02 in Figure 4b indicates the lower bound of the measurement, which corresponds to a thermal conductance (=$d \times \kappa_\parallel$) of ~10$^{-7}$ W m$^{-2}$ K$^{-1}$. The measurable lowest $\kappa_\parallel$ increases from 0.2 W m$^{-1}$ K$^{-1}$ to 6 W m$^{-1}$ K$^{-1}$ as the film thickness decreases from 1 μm to 10 nm. These bounds can be broadened when a wire with narrower width such as 50 nm is used.

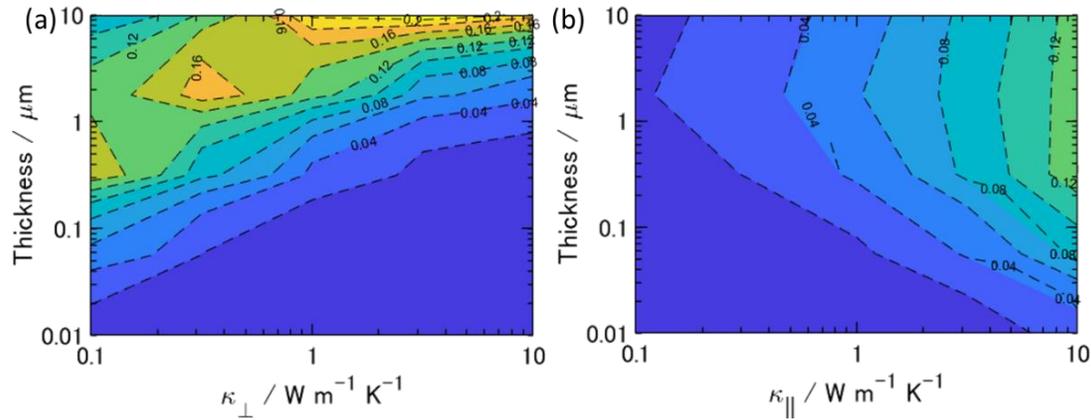

Figure 4 Contour of measurement sensitivity to (a)$\kappa_\perp$ and (b)$\kappa_\parallel$ of the model film sample with various thickness and $\kappa$.

## CONCLUSION

A novel bidirectional 3ω system with two different width wires to measure anisotropic thermal conductivity of organic thin film was developed. The metal wires of Ti / Pt with widths of 300 nm and 50 μm are patterned on quartz substrate beforehand and CNF, PMMA, and PEDOT:PSS thin films were prepared on the substrate. Cross-plane and in-plane thermal conductivity were measured in the sample with the 50-μm- and the 300-nm-wide wires, and this selective measurement was enabled by placing an aluminium layer on the samples in the measurement of cross-plane thermal conductivity and by suppressing the sensitivity to in-plane thermal conductivity. The in-plane thermal conductivity of 70-nm-thick film was successfully measured, revealing that high molecular ordering in the film can increase in-plane thermal conductivity of the thin film. Sensitivity calculations for various samples with different thermal conductivity and thickness clarified the applicable ranges, which suggests possibility for further improvement by narrowing the wire width.

## SUPPLEMENTARY MATERIAL

See supplementary material for additional information regarding the film preparation, schematic image and solution to the bidirectional heat transfer equation, material parameters, and signal with fitting curve for 3ω measurement.

## DATA AVAILAVILITY STATEMENT

The data that support the findings of this study are available from the corresponding author upon reasonable request.

## ACKNOWLEDGMENT

This work was supported in part by JSPS KAKENHI (Grant Nos. JP18J21726, 19H00744) and by JST CREST (Grant Nos. JPMJCR16Q5, JPMJCR19I2). This work was conducted at Takeda Sentanchi Supercleanroom, The University of Tokyo, supported by "Nanotechnology Platform Program" of the Ministry of Education, Culture, Sports, Science and Technology (MEXT), Japan, Grant Number JPMXP09F20UT0043.

# Anisotropic Thermal Conductivity Measurement of Organic Thin Film with Bidirectional 3ω Method

*Shingi Yamaguchi[1], Takuma Shiga [1], Shun Ishioka [2], Tsuguyuki Saito [2], Takashi Kodama [1], Junichiro Shiomi[1] \**

1. Department of Mechanical Engineering, The University of Tokyo, 7-3-1, Hongo, Bunkyo-ku, Tokyo 113-8656

2. Department of Biomaterial Sciences, Graduate School of Agricultural and Life Sciences, The University of Tokyo, 1-1-1 Yayoi, Bunkyo-ku, Tokyo 113-8657

・**Supplementary materials of bidirectional 3ω measurement**

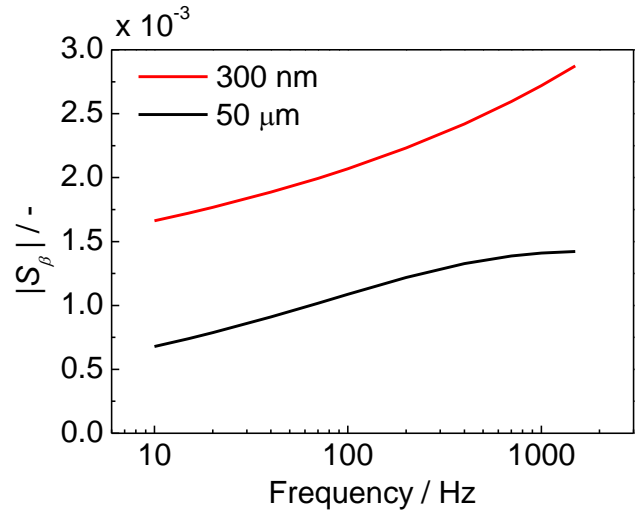

Figure S1 Measurement sensitivities to thermal boundary resistance (TBR) between $Al_2O_3$ and CNF, when the 50-μm-wide wire is used with Al deposition (black) and the 300-nm-wide wire is used without Al deposition (red).

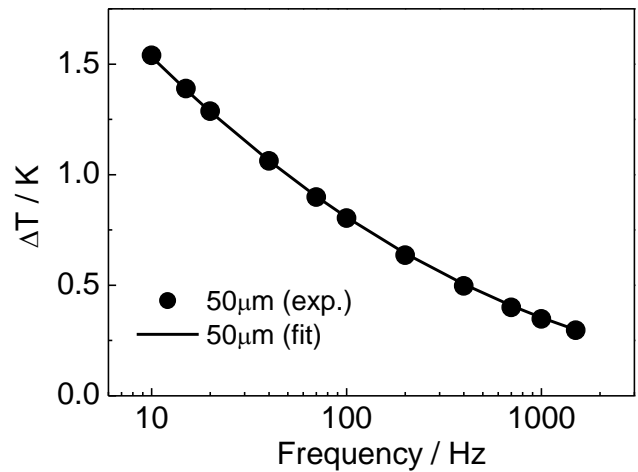

Figure S2 Representative temperature rise and its fitting curve for $\kappa_{Al}$ measurement, where 50-μm-wide wire were used with deposited 500 nm Al.

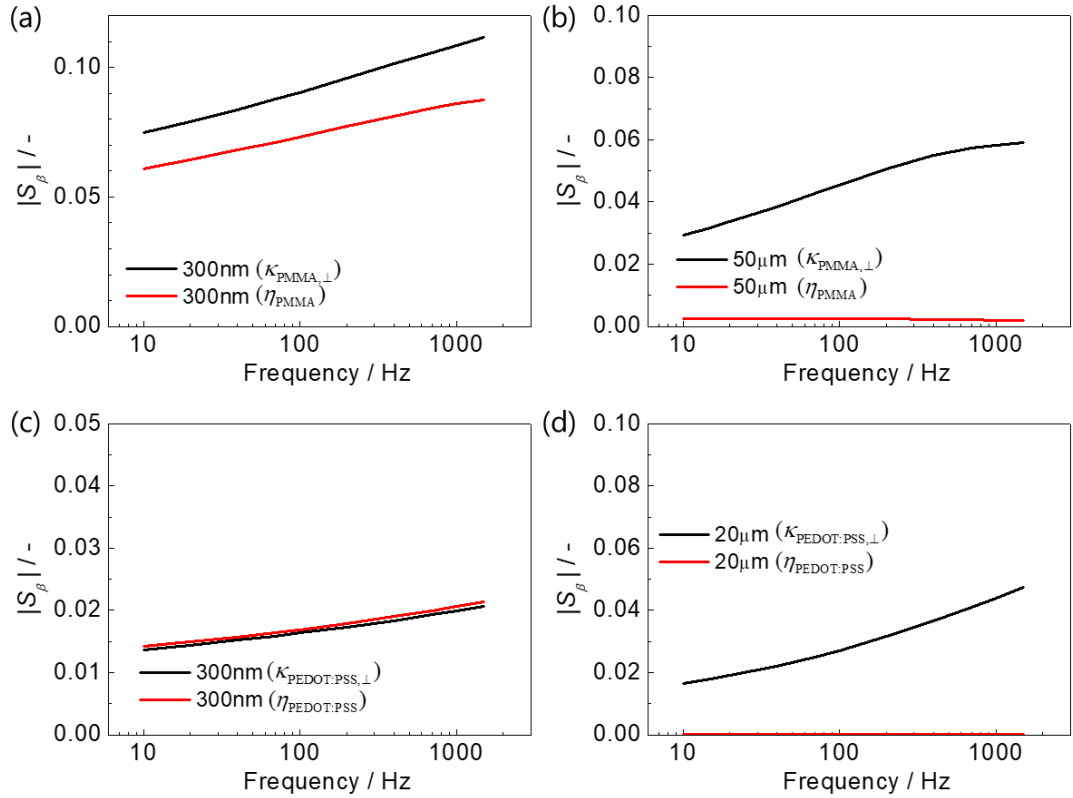

Figure S3 Sensitivities to $\kappa_\perp$ and $\eta$ of PMMA films when (a) 300-nm-wide wire is used without Al deposition and (b) 50-μm-wide wire is used with Al deposition, and sensitivities to $\kappa_\perp$ and $\eta$ of PEDOT:PSS films when (c) 300-nm-wide wire is used without Al deposition and (d) 20-μm-wide wire is used with Al deposition.

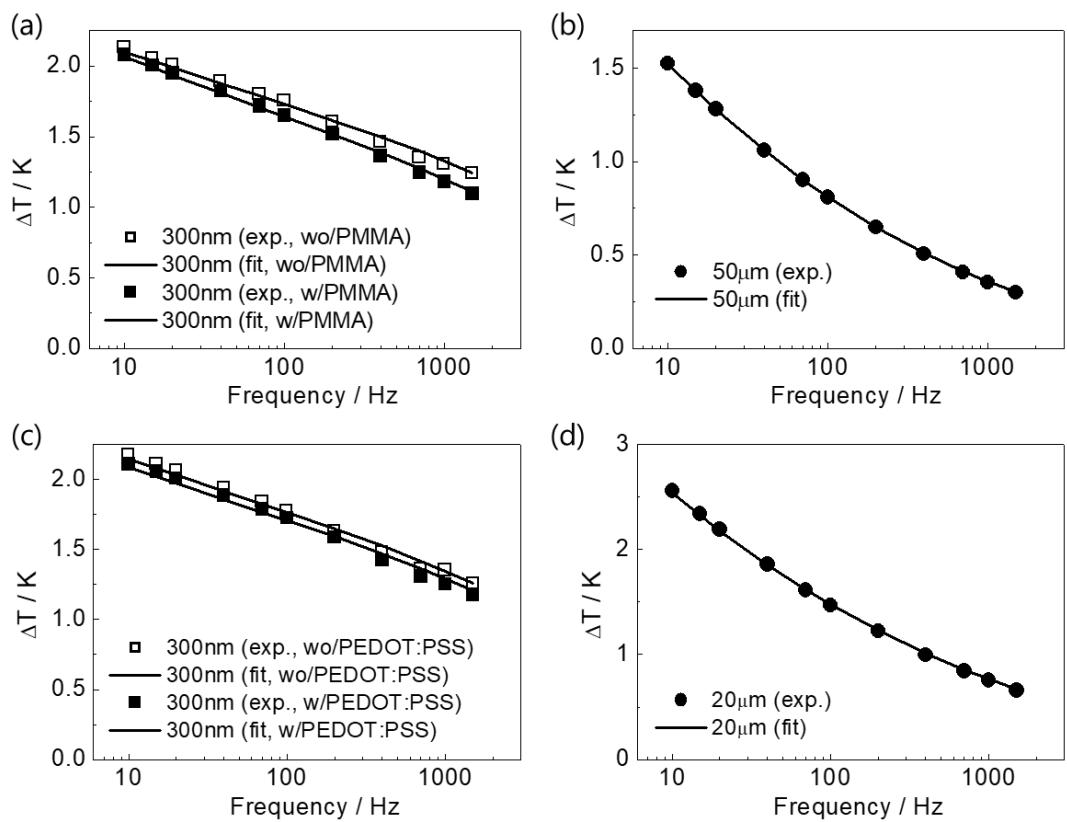

Figure S4 Representative temperature rise and its fitting curve for measurement of $\kappa_{\mathrm{PMMA}}$ and $\eta_{\mathrm{PMMA}}$ with (a) the 300-nm-wide wire and (b) 50-μm-wide wire, and $\kappa_{\mathrm{PEDOT:PSS,\perp}}$ and $\eta_{\mathrm{PEDOT:PSS}}$ with (c) the 300-nm-wide wire and (d) 20-μm-wide wire. Control measurement without PMMA or PEDOT:PSS is also shown in (a) and (c).

・**Thermal properties of each materials**

Table S1. Material parameters used for temperature calculation.

| | Heat Capacity (J/kg·K)* | Density (kg/m$^3$) |
|---|---|---|
| SiO$_2$ (substrate) | 760 | 2210** |
| Al$_2$O$_3$ | 780 | 3500$^2$ |
| Al | 900 | 2700$^3$ |
| CNF | 1273 | 1680*** |
| PMMA | 1610 | 1200$^3$ |
| PEDOT:PSS | 1980 | 1010** |

*Measured by differential scanning calorimetry (Netzsch, DSC 3500 Sirius).

**Provided by company (SiO$_2$ : TOSOH Quartz Materials, PEDOT:PSS : Sigma Aldrich).

***Measured from self-standing film.

・**Sample preparation**

PMMA film was prepared by spin coating 495 PMMA (MicroChem Corp.) on a device at 2000 rpm for 40 seconds. The device was then baked at 200 °C for 2 minutes to remove solvent.

For PEDOT:PSS film, mixed solution was firstly prepared. PEDOT:PSS (768642, Sigma-Aldrich) was mixed with 5 volume % of ethylene glycol, and then filtered by a syringe filter (0.45 µm pore-size PVDF membrane). Solutions were subsequently degassed in a vacuum chamber for 15 minutes. Films are prepared by spin coating at 4000 rpm for 30 seconds and by following baking at 130 °C for 15 minutes.

The CNF was prepared from a TEMPO-oxidized pulp, which was kindly provided by DKS Co. Ltd.. The TEMPO-oxidized pulp was suspended in distilled water at 0.5 weight %. The suspension was mechanically disintegrated into a CNF dispersion by passing it through a high-pressure water jet system (HJP-25005X, Sugino Machine Limited) two times. The CNF dispersion was diluted to 0.35 weight % with distilled water for preparation of a self-supporting CNF film. The 0.35 weight % CNF dispersion (80 mL) was poured into a 90-mm-diameter polystyrene petri dish and dried at 40 ºC and 80 % relative humidity. The resulting self-supporting CNF film was cut into a square with dimensions of 20 × 20 mm. For the CNF coated device, the CNF dispersion was concentrated to 1.0 weight % using a rotary evaporator at 40 ºC under reduced pressure. The 1.0 weight % CNF dispersion (100 mL) was dropped on the device and was dried in an oven at 40 ºC and 80 % relative humidity for a day.

・**Solution to the bidirectional heat transfer equation**

A general case for measurement system is shown in Figure S5. Here, j = 1 ~ N and j = 0, N+1 correspond to the stacked layers in the measurement device and outer matrix respectively. The heat diffusion equation for jth layer (j = 1 ~ N) is

$$D_{j,x}\frac{\partial^2 \Delta T_j(x,z)}{\partial x^2} + D_{j,z}\frac{\partial^2 \Delta T_j(x,z)}{\partial z^2} = \frac{\partial \Delta T_j(x,z)}{\partial t} \quad \cdots(S1)$$

where $\Delta T_j(x,z)$ is the temperature rise and $D_{j,x}$, $D_{j,z}$ is the thermal diffusivity in x and z direction, respectively. The left side of the jth layer is positioned at $z_j$ and the layer thickness is $L_j$. Here, the solution for Eq.S1 will be

$$\Delta T_j(x,z) = \Delta T_{j,\omega}(x,z)e^{2i\omega t} = \left[\int_{-\infty}^{\infty} \theta_j(z,k)e^{ikx}dk\right]e^{2i\omega t} \quad \cdots(S2)$$

and by substituting $\Delta T_j(x,z)$, Eq.S2 will be transformed to

$$\frac{\partial^2 \theta_j(z,k)}{\partial z^2} - (\eta_j k^2 + \frac{2i\omega}{D_{j,z}})\theta_j(z,k) = 0 \quad \cdots(S3)$$

where $\eta_j$ is the thermal anisotropy ($D_{j,x}/D_{j,z}$ or $\kappa_x/\kappa_z$). The general solution for Eq.S3 can be written as

$$\theta_j(z,k) = \theta_j^+(z) + \theta_j^-(z) = \theta_j^+ e^{u_j z} + \theta_j^- e^{-u_j z} \quad \cdots(S4)$$

$$\left(u_j = \sqrt{\eta_j k^2 + \frac{2i\omega}{D_{j,z}}} \quad \cdots(S5)\right)$$

To consider the heat transport in and between layers, the temperature will be represented as a vector $\Theta_j(z)$. Here, the temperature drop through the jth layer (between $z_j$ and $z_{j+1}$) can be expressed by a matrix $U(L_j)$ as shown below:

$$\Theta_j(z_{j+1}) = \begin{bmatrix} \theta_j^+(z_{j+1}) \\ \theta_j^-(z_{j+1}) \end{bmatrix} = \begin{bmatrix} e^{u_j L_j} & 0 \\ 0 & e^{-u_j L_j} \end{bmatrix} \begin{bmatrix} \theta_j^+(z_j) \\ \theta_j^-(z_j) \end{bmatrix} = U(L_j)\Theta_j(z_j) \quad \cdots(S6).$$

Also, the temperature drop between jth and (j+1)th layer is expressed by a matrix $\Gamma_{j,j+1}$:

$$\Theta_j(z_{j+1}^-) = \frac{1}{2\gamma_j}\begin{bmatrix} \gamma_j + \gamma_{j+1} - R_{j,j+1}\gamma_j\gamma_{j+1} & \gamma_j - \gamma_{j+1} + R_{j,j+1}\gamma_j\gamma_{j+1} \\ \gamma_j - \gamma_{j+1} - R_{j,j+1}\gamma_j\gamma_{j+1} & \gamma_j + \gamma_{j+1} + R_{j,j+1}\gamma_j\gamma_{j+1} \end{bmatrix}\Theta_{j+1}(z_{j+1}^+) = \Gamma_{j,j+1}\Theta_{j+1}(z_{j+1}^+) \quad \cdots(S7)$$

$$\left(\gamma_j = \kappa_j u_j \quad \cdots(S8)\right)$$

where $z_{j+1}^\pm$, and $R_{j,j+1}$ represent $z_{j+1} \pm 0$ and thermal boundary resistance (TBR), respectively. If the 3ω heater with a power density of $q$ is at the boundary between jth and (j+1)th layer, Eq.S7 would be

$$\Theta_j(z_{j+1}^-) = \Gamma_{j,j+1}\Theta_{j+1}(z_{j+1}^+) + \frac{q}{2\gamma_j}\begin{bmatrix} 1 \\ -1 \end{bmatrix} \quad \cdots(S9)$$

Here, TBR at $z_j$ is omitted. If the device is adiabatic at its bottom and surface ($z = z_1$ and $z_{N+1}$), $\Theta_0(z_1^-)$ and $\Theta_{N+1}(z_N^+)$ can be expressed as

$$\Theta_0(z_1^-) = \begin{bmatrix} \theta_0 \\ 0 \end{bmatrix} \quad \cdots(S10)$$

$$\Theta_{N+1}(z_{N+1}^+) = \begin{bmatrix} 0 \\ \theta_{N+1} \end{bmatrix} \quad \cdots(S11)$$

because Eq.S4 should converge at its limits of $z \rightarrow \pm\infty$. the Therefore, Eq.S9 can be transformed to

$$\theta_0 \mathbf{A} = \theta_{N+1}\mathbf{B} + \frac{q}{2\gamma_j}\begin{bmatrix}1\\-1\end{bmatrix} \quad \cdots \text{(S12)}$$

where

$$\mathbf{A} = \begin{bmatrix}A^+\\A^-\end{bmatrix} = U(L_j)\Gamma_{j,j-1}\cdots U(L_1)\Gamma_{1,0}\begin{bmatrix}1\\0\end{bmatrix} \quad \cdots \text{(S13)}$$

$$\mathbf{B} = \begin{bmatrix}B^+\\B^-\end{bmatrix} = \Gamma_{j,j+1}U(L_{j+1})\cdots U(L_N)\Gamma_{N,N+1}\begin{bmatrix}0\\1\end{bmatrix} \quad \cdots \text{(S14)}$$

From Eq.S12-S14, $\theta_0$ and $\theta_{N+1}$ will be

$$\theta_0 = \frac{q}{2\gamma_j}\frac{B^+ + B^-}{A^+B^- - A^-B^+} \quad \cdots \text{(S15)}$$

$$\theta_{N+1} = \frac{q}{2\gamma_j}\frac{A^+ + A^-}{A^+B^- - A^-B^+} \quad \cdots \text{(S16)}$$

and this will lead to the heater temperature ($\theta_j(z_j)$) as

$$\theta_j(z_j) = \frac{q}{2\gamma_j}\frac{(A^+ + A^-)(B^+ + B^-)}{A^+B^- - A^-B^+} \quad \cdots \text{(S17)}$$

By substituting $\theta_j(z_j)$ in Eq.S2 and Fourier transforming the heat density $q$, temperature rise will be

$$\Delta T_j(x,z) = \frac{P}{2\pi l}\int_0^\infty \frac{(A^+ + A^-)(B^+ + B^-)}{A^+B^- - A^-B^+}\frac{1}{\gamma_j}\frac{\sin^2(kb)}{(kb)^2}dk \quad \cdots \text{(S18)}$$

where $l$ is the heater length and $P$ is generating power calculated by $q \times l$.

In our case, bottom of the measurement device was kept isothermal by pasting thermal grease, and boundary condition of

$$\Theta_0(z_1^-) = \begin{bmatrix} \theta_0^+ \\ -\theta_0^+ \end{bmatrix} \cdots (S19)$$

was adopted instead of Eq.S10.

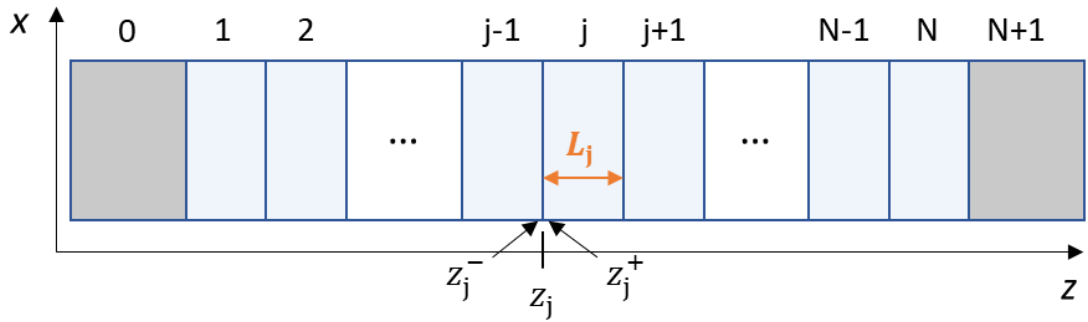

Figure S5 Schematic image of general multi-layer system with N components. jth layer with the thickness of $L_j$ positions at $z_j$. $0^{th}$ and $(N+1)$th layer represents outer matrix.

・**Origin of sensitivity difference with wide and narrow wires**

The sensitivity to $\kappa_{Al_2O_3}$ was larger when 300-nm-wide wire were used compared to 50-µm-wide wire case. Here, a model circuit with series of thermal resistance is considered to understand this sensitivity difference (Figure S6). $R_{Al_2O_3}$ and $R_{SiO_2}$ correspond to the thermal resistance of $Al_2O_3$ and $SiO_2$, and $\Delta T$ and $\Delta T_{surf}$ indicate the temperature rise of wire and $Al_2O_3$ surface. The calculated $\Delta T$ and $\Delta T_{surf}$ of each wire are shown in Table S2. There is no difference between $\Delta T$ and $\Delta T_{surf}$ in the 50-µm case, whereas 2 % temperature difference can be observed in the 300-nm case. This result shows that the relative $R_{Al_2O_3}$ gets increased when 300-nm is used, which leads to the increase of sensitivity to $\kappa_{Al_2O_3}$.

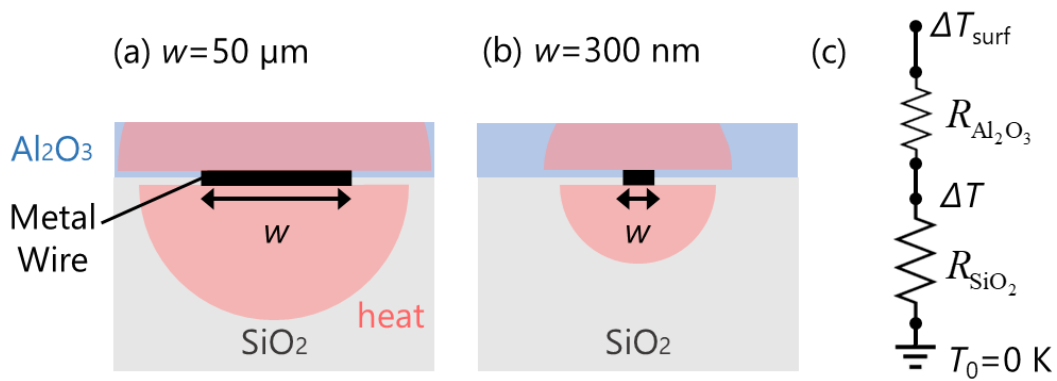

Figure S6 Cross-sectional image of heat conduction at (a) 50-µm- and (b) 300-nm-wide wire, and (c) corresponding thermal circuit.

Table S2 The temperature rise of wire and $Al_2O_3$ surface at frequency of 10 Hz.

|  | 50 µm | 300 nm |
| --- | --- | --- |
| $\Delta T$ / K | 1.71 | 2.59 |

| $\Delta T_{surf}$ / K | 1.71 | 2.55 |

・**Laser flash analysis**

The cross-plane thermal diffusivity of the CNF film was measured by laser flash analysis, using LFA 467 (Netzsch). Both sides of CNF self-standing film (thickness: 26 μm) was firstly covered with 100 nm Gold with resistive thermal evaporation, in order to suppress radiative heat conduction. The surface of the sample was then painted black with marker pens for the light absorption and detection.

The lamp voltage was set to 250 V, and pulse width was 3 μs. The representative signal and its fitting curve are shown in Figure S7.

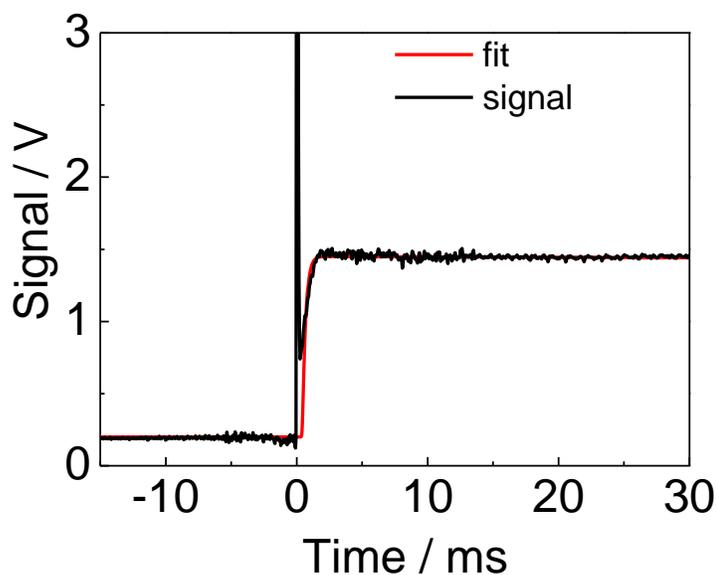

Figure S7 LFA signal and its fitting curve of CNF self-standing film.

・**T-type measurement**

In-plane $\kappa$ of CNF self-standing sample was measured by the T-type method. Figure S8 shows our experimental setup. A platinum wire with a diameter of 20 μm was suspended between two electrodes (copper blocks), and the electrical potential between them was measured while a constant current was applied. The sample was suspended between the center of the wire and the heat sink (another copper block), and the measured values of potential with and without the sample were fitted with an analytical expression derived from our theoretical model to extract the in-plane $\kappa$ of the sample. The details of analytical equation are described in the previous report[1].

The environmental temperature was set to 300 K, and was strictly controlled by Mercury iTC (Oxford Instruments). The temperature fluctuation for each measurement was kept below 0.1 K to suppress deviation of the signal.

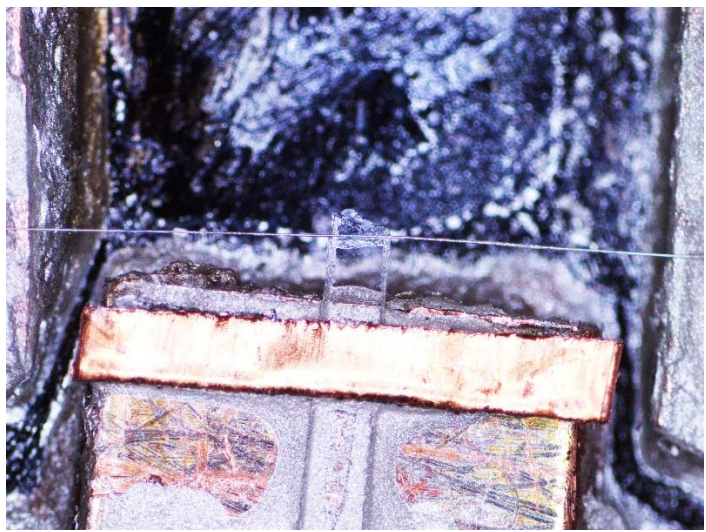

Figure S8 Experimental setup of T-type method.